\documentclass[manuscript]{aastex}

\usepackage{lscape}

\shorttitle{The First VLBI Image of 44~GHz methanol maser}
\shortauthors{Matsumoto et al.}

\begin{document}


\title{The First Very Long Baseline Interferometry Image of 44\,GHz Methanol Maser
    with the KVN and VERA Array (KaVA)}


\author{Naoko~Matsumoto\altaffilmark{1}, Tomoya~Hirota\altaffilmark{1, 2}, Koichiro~Sugiyama\altaffilmark{3}, 
 Kee-Tae~Kim\altaffilmark{4}, Mikyoung~Kim\altaffilmark{4}, Do-Young~Byun\altaffilmark{4},
Taehyun~Jung\altaffilmark{4}, James~O.~Chibueze\altaffilmark{5},  Mareki~Honma\altaffilmark{1, 2}, 
 Osamu~Kameya\altaffilmark{1}, Jongsoo~Kim\altaffilmark{4}, A-Ran~Lyo\altaffilmark{4}, Kazuhito~Motogi\altaffilmark{3}, 
Chungsik~Oh\altaffilmark{4}, Nagisa~Shino\altaffilmark{2}, Kazuyoshi~Sunada\altaffilmark{1}, 
Jaehan~Bae\altaffilmark{4, 6}, Hyunsoo~Chung\altaffilmark{4}, Moon-Hee~Chung\altaffilmark{4}, 
Se-Hyung~Cho\altaffilmark{4}, Myoung-Hee~Han\altaffilmark{4}, Seog-Tae~Han\altaffilmark{4}, Jung-Wook~Hwang\altaffilmark{4}, 
Do-Heung~Je\altaffilmark{4}, Takaaki~Jike\altaffilmark{1}, Dong-Kyu~Jung\altaffilmark{4}, Jin-seung~Jung\altaffilmark{4}, Ji-hyun~Kang\altaffilmark{4}, 
Jiman~Kang\altaffilmark{4}, Yong-Woo~Kang\altaffilmark{4}, Yukitoshi~Kan-ya\altaffilmark{1}, Noriyuki~Kawaguchi\altaffilmark{1, 2}, 
Bong~Gyu~Kim\altaffilmark{4}, Jaeheon~Kim\altaffilmark{4}, Hyo~Ryoung~Kim\altaffilmark{4}, Hyun-Goo~Kim\altaffilmark{4}, 
Hideyuki~Kobayashi\altaffilmark{1}, Yusuke~Kono\altaffilmark{1}, Tomoharu~Kurayama\altaffilmark{1, 7}, 
Changhoon~Lee\altaffilmark{4}, Jeong~Ae~Lee\altaffilmark{4}, 
Jeewon~Lee\altaffilmark{4, 8}, Jung-Won~Lee\altaffilmark{4}, Sang~Hyun~Lee\altaffilmark{4}, Sang-Sung~Lee\altaffilmark{4}, Young~Chol~Minh\altaffilmark{4}, 
Atsushi~Miyazaki\altaffilmark{4}, Se-Jin~Oh\altaffilmark{4}, Tomoaki~Oyama\altaffilmark{1}, Sun-youp~Park\altaffilmark{4}, Duk-Gyoo~Roh\altaffilmark{4}, 
Tetsuo~Sasao\altaffilmark{1, 4, 9}, Satoko~Sawada-Satoh\altaffilmark{1}, Katsunori~M.~Shibata\altaffilmark{1, 2}, Bong~Won~Sohn\altaffilmark{4}, Min-Gyu~Song\altaffilmark{4}, Yoshiaki~Tamura\altaffilmark{1}, Kiyoaki~Wajima\altaffilmark{10}, 
Seog-Oh~Wi\altaffilmark{4}, Jae-Hwan~Yeom\altaffilmark{4}, and Young~Joo~Yun\altaffilmark{4}}

\affil{\altaffilmark{1}Mizusawa VLBI Observatory, National Astronomical Observatory of Japan,
    2-21-1 Osawa, Mitaka, Tokyo 181-8588, Japan; naoko.matsumoto@nao.ac.jp}


\affil{\altaffilmark{2}Department of Astronomical Science, The Graduate University of Advanced Studies (SOKENDAI), 
\\ 2-21-1 Osawa, Mitaka, Tokyo 181-8588, Japan}

\affil{\altaffilmark{3}Graduate school of Science and Engineering, Yamaguchi University, 1677-1 Yoshida, Yamaguchi, Yamaguchi 753-8512, Japan}

\affil{\altaffilmark{4}Korea Astronomy and Space Science Institute, Daedeokdae-ro 776, Yuseong-gu, Daejeon 305-348, Korea}

\affil{\altaffilmark{5}East Asian ALMA Regional Center, National Astronomical Observatory of Japan,
2-21-1 Osawa, Mitaka, Tokyo 181-8588, Japan}

\affil{\altaffilmark{6}Department of Astronomy, University of Michigan, 500 Church Street, Ann Arbor, MI 48105, USA}

\affil{\altaffilmark{7}Center for Fundamental Education, Teikyo University of Science, 2525 Yatsusawa, Uenohara, Yamanashi 409-0193, Japan}

\affil{\altaffilmark{8}Department of Astronomy and Space Science, Kyung Hee University, Seocheon-Dong, Giheung-Gu, Yongin, 
Gyeonggi-Do 446-701, Korea}

\affil{\altaffilmark{9}Yaeyama Star Club, Ookawa, Ishigaki, Okinawa 904-0022, Japan}

\affil{\altaffilmark{10}Shanghai Astronomical Observatory, Chinese Academy of Sciences 80 Nandan Road, Xuhui District, Shanghai 200030, China}

\begin{abstract}
We have carried out the first very long baseline interferometry (VLBI) imaging of 44\,GHz class~\textsc{I} methanol maser 
(7$_{0}$$-$$6_{1}A^{+}$) associated with a millimeter core MM2 in a massive star-forming region \objectname{IRAS\,18151$-$1208} 
with KaVA (KVN and VERA Array), which is a newly combined array of  
KVN (Korean VLBI Network) and VERA (VLBI Exploration of Radio Astrometry).
We have succeeded in imaging compact maser features with a synthesized beam size of 
2.7\,milliarcseconds\,$\times$\,1.5\,milliarcseconds (mas).
These features are detected at a limited number of baselines within the length of shorter than 
$\approx$\,650\,km corresponding to 100\,M$\lambda$ in the $uv$-coverage.
The central velocity and the velocity width of the 44\,GHz methanol maser are consistent with those of 
the quiescent gas rather than the outflow traced by the SiO thermal line.
The minimum component size among the maser features is $\sim$\,5\,mas\,$\times$\,2\,mas, 
which corresponds to the linear size of $\sim$\,15\,AU\,$\times$\,6\,AU assuming a distance of 3\,kpc.
The brightness temperatures of these features range from $\sim\,3.5\,\times\,10^{8}$ to 
$1.0\,\times\,10^{10}$\,K, which are higher than estimated lower limit from a previous Very Large Array observation 
with the highest spatial resolution of $\sim$\,50\,mas.
The 44\,GHz class~\textsc{I} methanol maser in \objectname{IRAS\,18151$-$1208} is 
found to be associated with the MM2 core, which is thought to be less evolved than another 
millimeter core MM1 associated with the 6.7\,GHz class~\textsc{II} methanol maser.

\end{abstract}


\section{Introduction}
Methanol masers are known to be associated with star-forming regions. 
They are classified into two series of transitions called class~\textsc{I} and class~\textsc{II} \citep{Menten1991}.
The class~\textsc{I}~and~\textsc{II} methanol masers are considered to be excited by collisional 
\citep[e.g.,][]{Cragg1992} and radiative pumping mechanisms \citep[e.g.,][]{Sobolev1994, Cragg2005}, respectively.
In general, the class~\textsc{I} methanol masers are likely to be associated with interacting regions between outflows 
and dense ambient gases in both low- and high-mass star-forming regions \citep[e.g.,][]{Plambeck1990, Kalenskii2010}. 
In contrast, the class~\textsc{II} methanol masers are centrally concentrated around hot molecular cores, 
ultracompact\,(UC)\,H{\sc{ii}} regions, OH masers and near-IR sources only in high-mass star-forming regions \citep[e.g.,][]{Minier2003, Xu2008, Cyganowski2009, Breen2013}, 
and possibly associated with disk-like structures and outflows around massive young stellar objects \citep[e.g.,][]{Bartkiewicz2009, DeBuizer2003}.
The class~\textsc{I}~and~\textsc{II} methanol masers are complementary, 
and proper motion measurements of both maser lines are
 important to understand three-dimensional velocity structures of massive young stellar objects associated 
 with jet/outflow/circumstellar disk systems.
The class~\textsc{I}~and~\textsc{II} methanol masers also hold a possibility of evolutionary tracers, 
but the relationship between them is still unclear \citep[e.g.,][]{Ellingsen2007, Breen2010, Fontani2010}.

The $5_{1}$$-$$6_{0}A^{+}$ methanol maser at 6.7\,GHz is one of the brightest maser species, 
and classified as the class~\textsc{II}. 
The 6.7\,GHz methanol-maser emitting sources have been widely investigated with radio interferometers 
such as ATCA and Very Large Array \citep[VLA; e.g.,][]{Walsh1998}, and very long baseline interferometry \citep[VLBI; e.g.,][]{Sugiyama2008, Bartkiewicz2009}.
In contrast, the $7_{0}$$-$$6_{1}A^{+}$ methanol maser at 44\,GHz is representative of  the class~\textsc{I} methanol maser.
Recently, the 44\,GHz methanol masers have been extensively observed with single-dish radio telescopes 
and connected-element interferometers \citep[e.g.,][]{Kurtz2004}. 
 
However, there is a difficulty in observing the class~\textsc{I} methanol masers at high angular resolution.
Previous VLBI observations of the class~\textsc{I} methanol masers  
\citep[the $8_{0}$$-$$7_{1}A^{+}$ transition at 95.2\,GHz:][]{Lonsdale1998} have failed to detect any fringes 
for several strong class~\textsc{I} methanol maser sources. 
As the results of the VLA observations \citep[e.g.,][]{Slysh2002, Polushkin2009} and the negative results of \citet{Lonsdale1998}, 
spot sizes of the class~\textsc{I} methanol masers are expected to be between several milliarcseconds (mas) and $\sim$\,50~mas.
Because of such extended structures, it has been recognized that the class~\textsc{I} methanol masers are 
easily resolved out and hence, hardly detected with current VLBI instruments with a typical angular resolution of an order of $\sim$\,1\,mas. 
Thus, there was no report for successful imaging of the class~\textsc{I} methanol masers with VLBI to date. 

Recently, a new VLBI instrument, Korean VLBI Network (KVN) has been constructed, which consists of three 
21\,m radio telescopes \citep{Lee2014}. 
Since the maximum baseline length of KVN is $\sim$\,500\,km ($\theta_{\mathrm{beam}}\sim$\,3\,mas at 44\,GHz), 
it is the most suitable for conducting VLBI observations of the 44\,GHz class~\textsc{I} methanol masers (K.~T.~Kim et al., in preparation). 
To conduct high quality VLBI imaging, we extend the array by including four stations of VLBI Exploration of Radio 
Astrometry (VERA), which is another VLBI network in Japan. 

In this Letter, we report on the first result of our VLBI imaging observation of the 44\,GHz methanol maser 
line toward a star-forming region providing the highest angular resolution image. 
Our target source is one of the brightest 44\,GHz methanol maser sources, \objectname{IRAS\,18151$-$1208\,MM2}.
The distance of this source was estimated to be 3~kpc \citep{Sridharan2005}.
This source was detected in the course of a KVN single-dish survey with a total flux density of about 500\,Jy (K.~T.~Kim et al., in preparation), 
and hence, it is an ideal target to conduct high-resolution VLBI observations with KaVA (KVN and VERA Array), 
which is a newly combined array of KVN and VERA, with the longest baseline length of $\sim$\,2,300\,km \citep{Niinuma2014}.

\setcounter{footnote}{10}
\section{Observation and data reduction}
A VLBI observation was carried out on 2012 April 8, from UT\,17:10 to 23:53, with KaVA.
Our target source was the 44\,GHz methanol maser around \objectname{IRAS\,18151$-$1208\,MM2}. 
The phase tracking center position was 
$\alpha_{J2000.0} = 18^{\mathrm{h}}$17$^{\mathrm{m}}$50$\fs$1 and 
$\delta_{J2000.0} = -$12$\arcdeg$07$\arcmin$48$\arcsec$.
\objectname{NRAO\,530} was observed as a delay and bandpass calibrator.
A radio frequency of 44.069410\,GHz \citep[Tsunekawa et al. 1995;\footnote{http://www.sci.u-toyama.ac.jp/phys/4ken/atlas/}][]{Muller2004} is adopted 
in this Letter as the rest frequency of the CH$_{3}$OH $7_{0}$$-$$6_{1}A^{+}$ transition.
Left-hand circular polarization (LHCP) signals were recorded with SONY DIR1000 recorders at a rate of 128\,Mbit\,s$^{-1}$ at all the stations. 
A correlation process was carried out with the Mitaka\,FX correlator in NAOJ, Mitaka \citep{Chikada1991}. 
The resultant auto-correlation and cross-correlation spectra consist of 1024~spectral points with a frequency 
spacing of 31.25\,kHz ($\sim$\,0.21\,km\,s$^{-1}$) in two 16\,MHz bandwidth channels 
(1st~CH: 44.059$-$44.075~GHz, 2nd~CH 44.075$-$44.091~GHz in sky frequencies). 
In addition, both LHCP and RHCP (right-hand circular polarization) signals were recorded with Mark~5B recorders at a rate of 1024\,Mbit\,s$^{-1}$ at all the three KVN stations. 
They were correlated with the Distributed~FX \citep[DiFX;][2011]{Deller2007} software correlator in KASI, Daejeon \citep{Lee2014}.
Because we failed to record the data with the DIR1000 recorder at one of the KVN stations, Tamna, we instead employed the Mark\,5B/DiFX data 
for the Tamna baselines and combined with the DIR1000/Mitaka\,FX data for the other baselines. 
In this Letter, we only employed two 16\,MHz base-band signals of LHCP which are compatible with all the stations.

Data reduction was performed using the NRAO AIPS package by applying a standard procedure.
At first, the two datasets, correlated with DiFX and Mitaka\,FX, were calibrated separately as follows.
The delay and the delay-rate offset were calibrated using \objectname{NRAO\,530}. 
Bandpass responses were also calibrated using \objectname{NRAO\,530}. 
Amplitude calibrations were performed by a template method using the total-power spectra of 
the 44\,GHz methanol maser in \objectname{IRAS\,18151$-$1208\,MM2}.
Fringe-fitting was conducted for a reference maser component  in \objectname{IRAS\,18151$-$1208\,MM2} 
at a LSR (local standard of rest) velocity, $v_{\mathrm{LSR}}$, of 29.6\,km\,s$^{-1}$. 
Then, the datasets were combined after flagging a duplicated baseline in the data correlated 
with DiFX (the Yonsei-Ulsan baseline).
In the self-calibration of the reference maser component, visibility based model-fittings and self-calibrations were iteratively conducted 
with the DIFMAP software package provided by Caltech to detect faint components from the limited $uv$-coverage data. 
The self-calibration solutions were applied to the other velocity channels of the maser data.

\begin{figure}
\figurenum{1}
\label{vel-map}
\includegraphics[angle=0,scale=1.0]{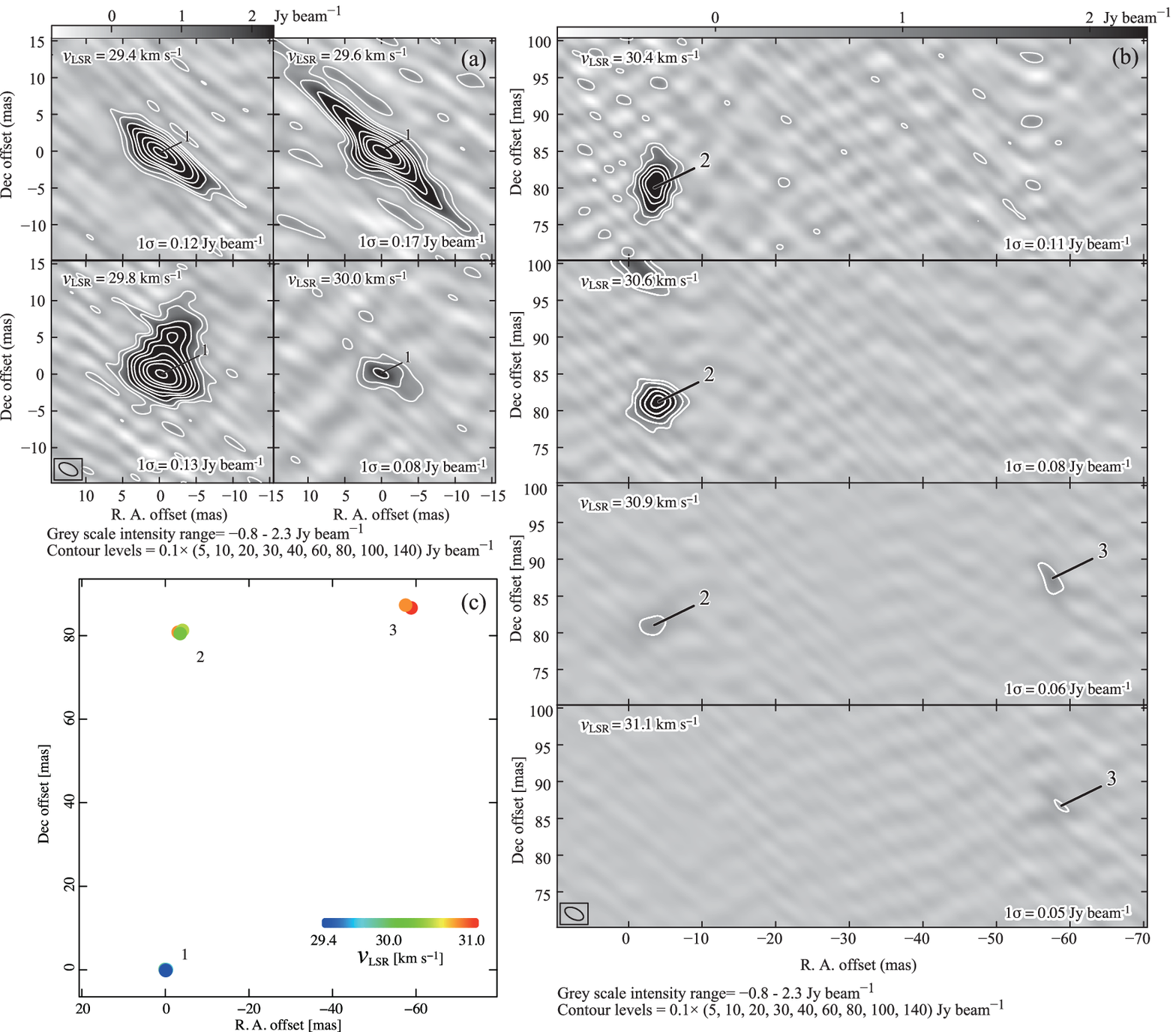}
\caption{(a) Multi-channel intensity maps of the feature~1, (b) same as (a) but for the features~2 and 3, and 
(c) a velocity map of the CH$_{3}$OH (7$_{0}$$-$$6_{1}A^{+}$) maser emission toward the IRAS\,18151$-$1208\,MM2.
The origin of these maps are $\alpha_{J2000.0} = 18^{\mathrm{h}}$17$^{\mathrm{m}}$50$\fs$0 and
 $\delta_{J2000.0} = -$12$\arcdeg$08$\arcmin$07$\arcsec$ with the errors of $\sim$\,20\arcsec.}
\end{figure}

Synthesis imaging and deconvolution were performed using DIFMAP with uniform weighting. 
In this procedure, cut-off signal-to-noise ratios (S/N) of 7$-$14 were used to avoid possible effects of side-lobes. 
These cut-off values depend on dynamic ranges of each channel map. 
The rms noise levels in the deconvolved maps are $\sim$\,50$-$200~mJy for \objectname{IRAS\,18151$-$1208\,MM2}.
The synthesized beam sizes (FWHM) and the position angle were 2.7\,mas\,$\times$\,1.5\,mas and 64$^{\circ}$, respectively. 
For all maser channels, maps were made with 8192\,pixel $\times$ 8192\,pixel with a pixel size of 0.1\,mas. 
Because of the insufficient velocity resolution and uncertainty in the clean process with our limited $uv$-coverage, maser structures 
in the images vary significantly from channel to channel (Figures\,\ref{vel-map} (a) and (b)). 
However, we regarded the three maser features identified in multi-channels 
within the synthesized beam size as real from a comparison between the visibility data and the convolution models. 
For example, the map of $v_{\mathrm{LSR}}$~=~29.8\,km\,s$^{-1}$ in Figure\,\ref{vel-map} (a) shows a double-peaked structure 
in a north-south direction while the northern component at ($-$2, 5) position was not detected in neighboring channels.
Therefore, this northern component is not considered in the following results and discussions.

Finally, the maser positions, fluxes, and sizes were derived by Gaussian fitting to each maser component in the identified three maser features 
using the AIPS task JMFIT (Table\,\ref{components}). 
To avoid influences from complicated structures of weak components, we employed image pixels with the intensity
 greater than the 90\,\% of the peak intensity in the fitting.
These maser components in the three maser features have S/N greater than 10 as seen in Table\,\ref{components} and Figures\,\ref{vel-map} (a) and (b). 
These peak intensities in the deconvolved maps correspond to 1.4$-$8.3 times the cut-off S/N in the imaging process.

\section{Results}
Figure~\ref{spectra} shows spectra of the 44\,GHz methanol maser line obtained from our observation. 
A velocity width of the 44\,GHz methanol maser line is around 3\,km\,s$^{-1}$. 
The weak red-shifted wing can be seen in the auto-correlation and cross-correlation spectra of the 44\,GHz methanol maser (Figure~\ref{spectra}).
The peak flux density of the 44\,GHz methanol maser is 379\,$\pm$\,22\,Jy ($v_{\mathrm{peak}}$\,=\,30.0\,km\,s$^{-1}$) in the auto-correlation spectrum.
However, the maximum value of the cross-correlation spectrum is 102\,$\pm$\,1\,Jy ($v_{\mathrm{peak}}$\,=\,29.5\,km\,s$^{-1}$) in 
Yonsei-Ulsan baseline at UT\,18:46\,-\,18:47 corresponding to $u$\,$\sim$\,18\,M$\lambda$ and $v$\,$\sim$\,19\,M$\lambda$.
Furthermore,  except UT\,17\,h$-$21\,h of Yonsei\,-\,Ulsan baseline, peak flux densities of all the baselines are lower than $\sim$\,20\,Jy.
Thus, 27\,\% of the emission can be recovered with our VLBI images and the remaining 73\,\% is completely resolved out. 
This means that the 44\,GHz methanol maser emission is not a compact point source but is dominated by a spatially extended 
structure. 

Such an extended structure of each feature is also shown in a $uv$-distance plot (Figure~\ref{uv-amp}). 
This plot shows a variation of the visibility amplitude with the $uv$-distance for the systemic velocity of 29.8\,km\,s$^{-1}$. 
There is a decreasing trend of the amplitude within $\sim$\,27\,M$\lambda$. 
The data points between $\sim$\,100$-$330\,M$\lambda$ could not be detected in the fringe-fitting process.
From Gaussian fitting to this plot along with the flux of auto-correlation spectrum, this trend suggests an extended maser component with a size of $\sim$\,4\,mas. 
In addition, an increasing trend of the amplitude around 20\,M$\lambda$ suggests an elongated maser structure.
The visibility data with the $uv$-lengths shorter than 20\,M$\lambda$ corresponds to the Yonsei-Ulsan baseline data at UT\,17:16\,-\,18:48.
In contrast, visibility amplitudes with the baseline lengths longer than 30~M$\lambda$ have almost constant values of $\sim$\,15\,Jy, 
suggesting a spatially compact structure.

Figure~\ref{vel-map}~(c) shows a spatial and velocity distribution of the 44\,GHz methanol maser components.
We found three maser features at different LSR velocities, denoted as 1$-$3. 
The absolute position of the maser component with $v_{\mathrm{LSR}}$\,=\,29.4\,km\,s$^{-1}$ 
at (0, 0) position is derived from a fringe-rate mapping using the AIPS task FRMAP to be 
$\alpha_{J2000.0} = 18^{\mathrm{h}}$17$^{\mathrm{m}}$49$\fs$95, 
and $\delta_{J2000.0} = -$12$\arcdeg$08$\arcmin$06$\farcs$5. 
Thus, the 44\,GHz methanol maser components are separated from the MM2 core position \citep{Beuther2002a}
by $\Delta\alpha$\,$\sim$\,0.4\arcsec and $\Delta\delta$\,$\sim$\,11.5\arcsec.

As mentioned in Section\,1, the 44\,GHz methanol maser toward the \objectname{IRAS\,18151$-$1208\,MM2}
 was detected in the course of KVN single-dish survey observations (K.\,T.\,Kim et al. in preparation). 
The single-dish observation was carried out toward the MM1 position, but the bright maser emission was 
found to be offset toward the southwest direction. According to the five-point cross-scan,
the maser position was identified as the MM2 position with the accuracy of $\sim$\,5\arcsec. 
In our VLBI observation, we adopted the coordinates obtained from the single-dish survey. 
However, the coordinates obtained from the fringe-rate mapping was different from the single-dish result 
by $\sim$\,19\arcsec in a north-south direction, which is larger than the position error. 
The difference would be caused by a larger uncertainty in the fringe-rate mapping probably due to our insufficient $uv$-coverage  
which made the lines in the fringe-rate map roughly lie north-south directions. 
Even if the result of the fringe-rate mapping has the error of around 20\arcsec, 
the maser components detected in our observation are considered to belong to the MM2 region \citep[see Figure~1 of][]{Beuther2002a}.

The radial velocities of these components range from $v_{\mathrm{LSR}} = 29.4$ to 31.1\,km\,s$^{-1}$.
They are spread over an area of $\sim$\,60\,mas\,$\times$\,90\,mas or $\sim$\,180\,AU\,$\times$\,270\,AU 
at an assumed distance of 3\,kpc \citep{Sridharan2005}. 
We cannot see significant velocity structure indicative of outflow, infall, or rotational motions. 

The channel maps are shown in Figures~\ref{vel-map} (a) and (b) for the 44\,GHz methanol maser features.
These maser structures are clearly more extended than the beam size. 
Their detailed properties are summarized in Table~\ref{components}.
The minimum size among the maser features is 4.98\,($\pm$\,0.32)\,mas\,$\times$\,2.16\,($\pm$0.14)\,mas at $v_{\mathrm{LSR}}$ of 30.9\,km\,s$^{-1}$, 
and the maximum brightness temperature is estimated to be $1.0 \times 10^{10}$~K at $v_{\mathrm{LSR}}$ of 29.6\,km\,s$^{-1}$ with 
an uncertainty in the absolute amplitude calibration around a few tens percent in addition to the fitting errors in Table~\ref{components}.

\section{Discussions}
We have succeeded in detecting the 44\,GHz methanol maser features around the IRAS\,18151$-$1208 MM2 using newly organized KaVA. 
This is the first VLBI image of the class~\textsc{I} methanol maser line at 44\,GHz with the highest angular resolution of about 2\,mas. 
Thanks to the high angular resolution, our results suggest that the 44\,GHz methanol masers also have compact components 
with the sizes of a few tens of AU that are detectable by VLBI observations. 
With further observations of the class~\textsc{I} methanol masers, 
it will be possible to provide three-dimensional velocity maps including radial velocities and proper motions as well as 
useful information to reveal the maser pumping mechanisms by comparing positions of multiple methanol maser 
lines at high spatial resolutions comparable with their spot sizes. 

Previously, the highest resolution maps of the 44\,GHz methanol masers were reported by \citet{Polushkin2009} 
for \objectname{DR21(OH)} using VLA with the resolution of 50\,mas. 
They derived the linear sizes of the maser components to be 30$-$480\,AU.
\citet{Slysh2002} also reported an upper limit of the maser component size of typically 50\,mas, 
and their linear sizes are less than several hundred AU for five sources 
with VLA observation with an angular resolution of $\sim$\,100\,mas.
They showed the lower limit of the brightness temperature of 10$^8$\,K for the strongest masers.

In our VLBI images, the minimum size among the maser features is estimated to be 2$-$4\,mas, 
and the maximum brightness temperature is estimated to be $\sim$\,10$^{10}$\,K. 
This brightness temperature is much higher by two to four orders of magnitude than those estimated 
by previous lower resolution observations with VLA \citep[e.g.,][]{Kogan1998, Slysh2002}.
This is the first case determining the brightness temperature for the class~\textsc{I} methanol masers with a sufficiently high spatial resolution, 
and consistent with the previous suggestions \citep[e.g.,][]{Kogan1998, Slysh2002}.

\begin{figure}
\figurenum{2}
\label{spectra}
\includegraphics[angle=0,scale=0.9]{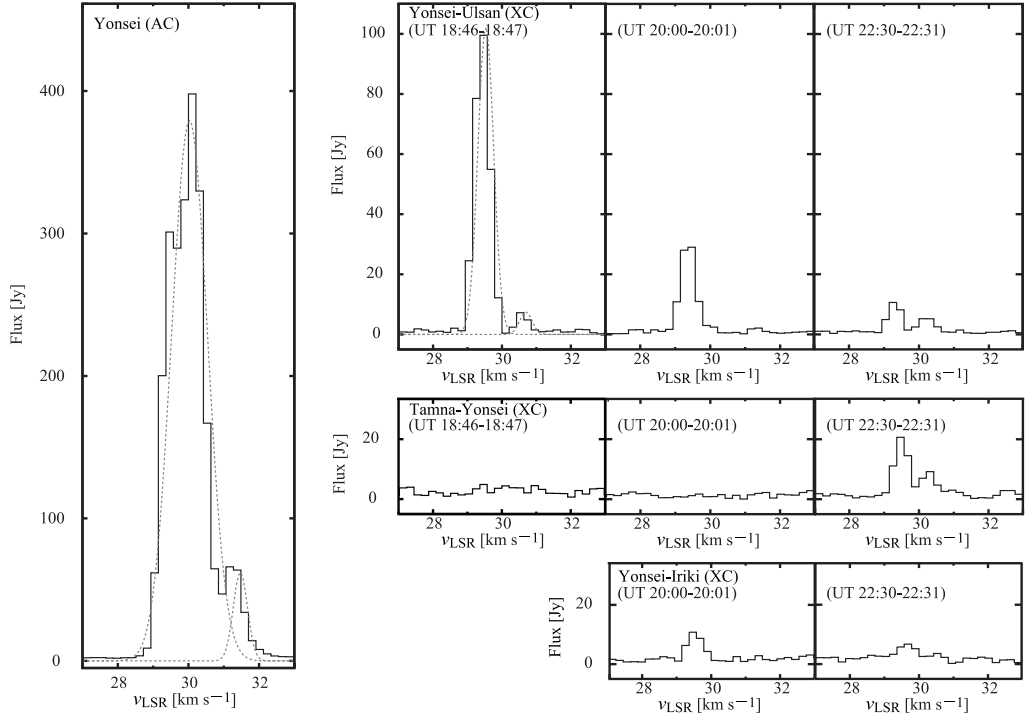}
\caption{Spectra of 44\,GHz methanol maser lines of auto-/cross-correlation (AC/XC). 
Dashed lines are Gaussian components obtained by the method of least squares.}
\end{figure}

\begin{figure}
\figurenum{3}
\label{uv-amp}
\includegraphics[angle=0,scale=1.0]{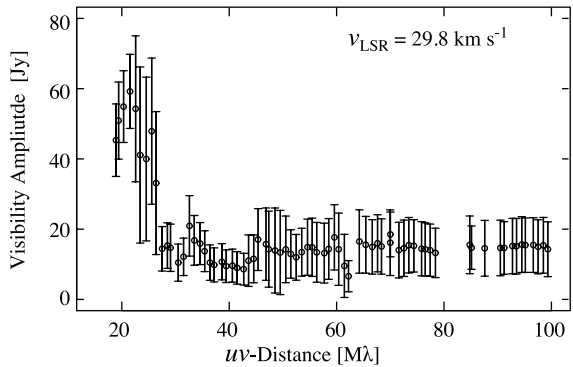}
\caption{The $uv$-distance plots for visibility amplitude of a CH$_{3}$OH (7$_{0}$$-$$6_{1}A^{+}$) maser component toward 
the IRAS\,18151$-$1208\,MM2 at a $v_{\mathrm{LSR}}$ of 29.8\,km\,s$^{-1}$ .}
\end{figure}

As seen in Figure~\ref{spectra}, 
the central velocity of the 44\,GHz methanol maser agrees well with the ambient molecular gas traced by the H$^{13}$CO$^{+}$ thermal line \citep[see Figure~1 of][]{Sakai2010}. 
Although it is proposed that the 44\,GHz methanol maser is excited in the shocked molecular gas, 
both central velocity and its velocity range are not well correlated with the outflows traced by the SiO thermal line \citep[see Figure~1 of][]{Sakai2010}.
It is consistent with previous interpretations for other sources that the class~\textsc{I} methanol maser at 95\,GHz traces 
intermediate gas around interaction regions with protostellar outflows and ambient quiescent molecular gas \citep{Plambeck1990}.

In the \objectname{IRAS\,18151$-$1208} region, there are four millimeter massive cores labeled MM1 to MM4 \citep{Beuther2002a} and two of them, 
MM1 and MM2, are associated with maser emissions \citep{Beuther2002b}. 
The 6.7\,GHz class~\textsc{II} methanol maser is detected toward MM1 \citep[e.g.,][]{Beuther2002b}, 
and the 22\,GHz H$_{2}$O maser is detected toward both MM1 \citep{Valdettaro2001} 
and MM2 \citep{Beuther2002b}. 
These water/methanol maser lines and CO outflows \citep{Marseille2008} are signposts of young stellar objects.

MM2 is considered as a chemically and dynamically less-evolved massive core than MM1 \citep[e.g.,][]{Marseille2008, Sakai2010, Chen2011a, Ragan2012}. 
MM1 is the birthplace of a massive young stellar object, possibly a pre-UC\,H{\sc{ii}} region \citep[e.g.,][]{Campbell2008, Marseille2008}. 
In fact, around the MM1 position, radio continuum emissions are detected at $\lambda \sim$\,1.3\,cm \citep{Beuther2009, Hofner2011} and 
3.6\,cm \citep{Carral1999}, which are interpreted as shock-ionized flows/jets rather than UC\,H{\sc{ii}} region \citep{Hofner2011}. 
In contrast, there is no detection of the 6.7\,GHz methanol maser or radio continuum at cm-wavelength around the MM2 position. 

Based on infrared observations, MM1 is also identified as a high-mass protostellar object with a point source of 
$MSX$~8\,$\mu$m and $Spitzer$~24\,$\mu$m emissions \citep{Sakai2012}, while MM2 is an $MSX$ dark source and 
associated with the $Spitzer$~24\,$\mu$m emission \citep{Sakai2012}. 
$Herschel$ data also show that MM1 and MM2 are associated with far-infrared point sources in 70$-$500\,$\mu$m images, 
and MM1 is brighter than MM2 \citep{Ragan2012}. 
From the $Herschel$ data, the masses of MM1 and MM2 cores are estimated to be 106\,$M_{\odot}$ and 81\,$M_{\odot}$, respectively \citep{Ragan2012}. 

All these characteristics suggest that the evolutionary stage of MM2 could be earlier than MM1.
While we cannot rule out a possibility that MM2 does not host any massive young stellar objects judging from a non-detection of 
the 6.7\,GHz class~\textsc{II} methanol maser, it is a potential site of massive star formation with a sufficient mass of 81\,$M_{\odot}$. 
If this is the case, our result suggests that the 44\,GHz class~\textsc{I} methanol maser tends to appear in an earlier evolutionary phase 
than that traced by the 6.7\,GHz class~\textsc{II} methanol maser in the \objectname{IRAS\,18151$-$1208} region. 
It is consistently explained by one of the currently proposed scenarios of the evolutionary phase of 
massive star-forming regions that the class~\textsc{I} methanol masers appear in a younger phase than 
the class~\textsc{II} methanol masers \citep{Ellingsen2007}. 
However, other observational studies claimed another trend in which the class~\textsc{I} methanol masers 
may be associated with more than one evolutionary stage (i.e., hot core and UC\,H{\sc{ii}}) 
during formation of massive young stellar objects \citep{Voronkov2010, Chen2011b, Cyganowski2012}. 
One possible reason for such a discrepancy would be insufficient spatial resolutions employed 
in some of the survey observations with single-dish telescopes, which could be affected by the complicated 
structure of massive star-forming regions as found in our target source \objectname{IRAS\,18151$-$1208}. 
Thus, further high-angular resolution observations of a much larger sample of methanol maser sources 
are required to discuss the evolutionary stage of massive young stellar objects with maser species. 
VLBI observations of the 44\,GHz methanol masers will be powerful tools to provide their accurate positions and proper motions, 
which are crucial to identifying the powering source of masers.

\acknowledgments

We are grateful to all staff members and students at the KVN and VERA who helped to operate the array and to correlate the data.
The KVN is a facility operated by the Korea Astronomy and Space Science Institute. 
VERA is a facility operated by National Astronomical Observatory of Japan in collaboration with Japanese universities. 
We also appreciate Dr. T.~Sakai and Dr. S.~Kameno for useful comments. 

{\it Facilities:} \facility{VERA}, \facility{KVN}

\clearpage


\clearpage

\begin{landscape}
\begin{table}
\tablenum{1}
\begin{center}
\caption{List of Identified maser components\label{components}}
\begin{tabular}{ccr@{ (} lr@{ (} lr@{ (} lr@{ (} lr@{ (} lr@{ (} l}
\tableline\tableline
Feature & $v_{\mathrm{LSR}}$ & \multicolumn{2}{c}{$\Delta \alpha$} & \multicolumn{2}{c}{$\Delta \delta$} & \multicolumn{2}{c}{Peak Intensity} & \multicolumn{2}{c}{Component size} & \multicolumn{2}{c}{P. A.} & \multicolumn{2}{c}{$T_{\mathrm{peak}}$}\\
 ID &  [km\,s$^{-1}$] & \multicolumn{2}{c}{[mas]} & \multicolumn{2}{c}{[mas]} & \multicolumn{2}{c}{[Jy beam$^{-1}$]} & \multicolumn{2}{c}{[mas]} & \multicolumn{2}{c}{[$^{\circ}$]} & \multicolumn{2}{c}{[K$\times$10$^{8}$]}\\
\tableline
1 & 29.4 &   $-$0.10&3) & $-$0.19&2) & 10.74&12) & 7.01 $\times$ 2.51&8, 3) &   55.1&4) & 67.5&8)\\ 
1 & 29.6 &   $-$0.07&2) & $-$0.07&2) & 16.33&17) & 6.03 $\times$ 2.52&6, 3) &   59.9&4) & 103&1)\\
1 & 29.8 &   $-$0.27&2) &	     0.03&1) & 14.90&12) & 5.61 $\times$ 3.13&5, 3) &   65.8&5) & 93.7&8)\\
1 & 30.0 &        0.09&8) &	     0.12&5) &   2.30&9) & 5.10 $\times$ 2.29&19, 9) &   63.4&17) & 14.4&5)\\
2 & 30.4 &   $-$3.53&3) &   80.52&5) & 5.43&10) & 6.39 $\times$ 3.73&12, 7) & 162.3&13) & 34.2&6)\\
2 & 30.6 &   $-$4.06&3) &	   81.25&3) &   4.49&8) & 4.96 $\times$ 3.91&8, 7) &   120.8&29) & 28.2&5)\\
2 & 30.9 &   $-$3.06&14) &   80.82&10) &   0.74&6) & 4.30 $\times$ 3.15&32, 24) & 82.8&96) & 4.68&35)\\
3 & 30.9 & $-$57.50&9) &    87.32&12) &   0.87&6) & 4.98 $\times$ 2.16&32, 14) &   32.5&28) & 5.46&35)\\
3 & 31.1 & $-$58.82&21) &  86.62&18) &   0.55&1) & 6.70 $\times$ 2.14&62, 20) &   50.3&27) & 3.45&3)\\
\tableline
\end{tabular}
\tablecomments{The numbers in parentheses represent the standard deviation in the Gaussian fit in units of the last significant digits.}
\end{center}
\end{table}
\end{landscape}

\end{document}